\documentclass[preprint,prl,aps,floats,showpacs]{revtex4}
\usepackage{graphicx}
\usepackage{amsmath}
\usepackage{amssymb}
\usepackage{wrapfig}
\begin{document}
\title{Gross features of finite nuclei at finite temperatures}
\author{S.M.M. Coelho\footnote{E-mail:sergio@up.ac.za}, C. Zander\footnote{E-mail:cz@up.ac.za} and H.G. Miller\footnote{E-mail:hmiller@maple.up.ac.za}}
\affiliation{Department of Physics, University of Pretoria, Pretoria 0002, South
Africa}
\begin{abstract}
A simple expression is obtained for the low temperature behavior of the
energy and entropy  of finite nuclei for $20\leq A\leq 250$. The
dependence on $A$ of these quantities is for the most part due to the presence of
the asymmetry energy. 
\end{abstract}
\maketitle
In nuclei the  nuclear force between the nucleons is short-ranged which leads to
saturation of the binding energy, $E(A,Z)$, per nucleon.  Empirically it is known
that for the stable isotopes 
\begin{equation}
\frac{ E(A,Z)}{A}\approx \alpha
\end{equation}
where $\alpha\approx 8$ MeV  and $A\geq 20$. This fundamental gross property of
nuclei is well accounted for by the semi-empirical mass
formula \cite{BB35,VW35,S61} which provides a simple
parametrization of the binding energy per nucleon for all known nuclei. 
Qualitatively it is also consistent with the simple Fermi gas model
prediction \cite{LP80,P84}
\begin{equation}
 \frac{\mathcal{ E}(A)}{A} =\frac{3}{5}\epsilon_f
\end{equation}
where $\mathcal{E}$ is the ground state energy and $\epsilon_f$ is the Fermi
energy which is constant as long as the particle density remains constant. It is
interesting to note that at low temperatures ($T<T_f$) the excitation energy in
the Fermi gas model is given by \cite{LP80,M91}
\begin{eqnarray}
 \frac{\mathcal{E}(A,T)}{A} &=&
\frac{3}{5}\epsilon_f+\frac{\pi^2}{4}\frac{T^2}{\epsilon_f} \hspace{3mm}
\text{or}\\
   &=& a+ b T^2 \label{et}
\end{eqnarray}
where $a$ and $b$ are constant again as long as the particle density remains
constant. Although the Fermi gas model may be an oversimplified model, nonetheless it underscores the relevance of independent particle (or
quasi-particle) methods in nuclear structure physics. Mean field methods have
been used throughout the periodic table, both  at zero temperature and at finite
but low temperatures, and typically yield a $T^2$ behaviour of the energy
density. This suggests that perhaps at 
low but finite temperatures a simple scaling relation might also exist for the
excitation energy, $\mathcal{E}(A,T)$, of finite nuclei.

In order to test the validity of equation (\ref{et}) we have made use of  a
finite temperature extension of the semi-empirical mass formula \cite{DHMMT93}.
The following  form for the
temperature dependent binding energy \cite{S61,DeB64} has been assumed
\begin{eqnarray}
E(A,Z,T) &=& \overbrace{\alpha(T)A}^1 +\overbrace{\beta(T)A^{\frac {2} {3}}}^2
+\overbrace{(\gamma(T)
-\frac {\eta(T)} {A^\frac {1} {3}}) (\frac {4t^2_{\zeta} +
4|t_{\zeta}|} {A})}^3 \nonumber \\
& & \mbox{}+\underbrace{\kappa(T) \frac {Z^2}{A^\frac {1} {3}} (1-\frac{0. 7636}
{Z^\frac{2}{3}}
-\frac{2.29\,\kappa^2(T)}{(0.8076)^2 A^{\frac{2}{3}}})}_4+\underbrace{\delta (T)
f(A,Z) A^{- \frac {3}{4}}}_5 \label{binding energy}
\end{eqnarray}
where  $A=N+Z$,  $t_{\zeta} = \frac{1}{2}(Z-N)$ and $f(A,Z)
=(-1,0,+1)$ for (even-even, even-odd, odd-odd) nuclei. Here 1 is the
volume energy, 2 is the surface energy, 3 is the asymmetry energy, 4 is the
Coulomb energy and 5 is the pairing energy contribution to temperature
dependent binding energy.
In this parametrization  the temperature
dependence of the contributions to the Coulomb energy term which
arise from
exchange and surface effects \cite {S61,M59} was ignored. Also,  no
attempt  has
been made to
include shell effects in the finite temperature expression.

The excitation energy per particle is given by
\begin{equation}
\frac{\mathcal{E}(A,Z,T)}{A} = \frac{E(A,Z,T) - E(A,Z,0)}{A}.
\label{eq:exc1}
\end{equation}
\noindent At $T=0$ the coefficients are given by \cite{S61}
$\alpha(0) = -16.11 $ MeV,
$\beta(0) = 20.21 $ MeV,
$\gamma(0) = 20.65 $ MeV,
$\eta(0) = 48.00 $ MeV, and
$\kappa(0) = 0.8076 $ MeV obtained from a fit to the experimental
nuclear
ground state energies of 488 odd mass nuclei.
The $T=0$ coefficient for the pairing term is taken as
$\delta(0)=33.0 $ MeV \cite{DeB64}.

To obtain the temperature dependence of the coefficients,
the available experimental information about the excited states of
nuclei
throughout the periodic table was used to determine the partition function
of each nucleus in the canonical ensemble
\begin{equation}
 \mathcal{Z}(A,Z,T) = \sum^n_i {g_i \exp(-\beta E_i)} +\int^{
E_{max}}_{E_n} dE \
g_{A,Z}(E) \exp(-\beta E) \label{eq:part}
\end{equation}
\noindent where $g_i=2j_i + 1$ is the spin degeneracy factor and $E_i$ the
excitation energy of the $i$th state of the nucleus, and $\beta
=1/T$. The sum
in the first
term of equation (\ref{eq:part}) runs over the experimentally
measured
(discrete) excited states.

Since the experimentally known spectrum in most cases is only sufficient
to allow
the accurate determination of ${\mathcal Z}$ for very low temperatures ($T
\ll
1$~MeV), it is necessary to supplement the experimentally known
spectrum with an appropriate
approximation to the continuum $g_{A,Z}(E)$. For this purpose, the fits
obtained in \cite{GC65} were used. For sufficiently large energies, the usual
Fermi
gas expression for the total density of states ($i.e.$ including the
spin
degeneracy) is used:
\begin{equation}
g_{A,Z}(E) =  \frac{\sqrt{\pi}}{12} \frac{\exp{(2 \sqrt{a_{A,Z}U})}}
{a^{1/4}_{A,Z} U^{5/4}}. \label{eq:gfg}
\end{equation}
Here $a_{A,Z}$ is the level density parameter and $U = E - P(N) -
P(Z)$,
where $P(N)$ and $P(Z)$ are the pairing corrections for neutron
number $N$
and proton number $Z$ respectively \cite{GC65}. This parametrization
is
obtained by means of a saddle point approximation, and is probably
only valid up
to $T \approx 6$ MeV. However, in the
region up to $T
\approx 4$ MeV, this parametrization is probably  acceptable.

At lower energies, a suitable
fit to the nuclear energy level density can be obtained to the form
\begin{equation}
g_{A,Z}(E) = \frac{\sqrt{2\pi}\sigma}{\tau}\,\, \frac{\exp{(E-E_0)}}{\tau},
\label{eq:gexp}
\end{equation}
with $\sigma$ the spin-dependence parameter. Values for the
parameters
$a_{A,Z},\,\tau,\,E_0$ and $\sigma$, as well as the respective
regions where
(\ref{eq:gfg}) and (\ref{eq:gexp}) should be used, for a large
number of nuclei can be found in \cite{GC65}.

The nuclei used to determine the aforementioned coefficients
can therefore be divided into three
groups: Nuclei where sufficient discrete states are known to allow the
use of
the discrete spectrum at low energies and the Fermi gas expression
(\ref{eq:gfg}) at higher energies. Nuclei where the discrete spectrum does not
extend high enough
for
(\ref{eq:gfg}) to be valid. For these nuclei, the discrete spectrum
is used
for low excitation energies, followed by the exponential form
(\ref{eq:gexp}) for
intermediate energies and finally the Fermi gas expression (\ref{eq:gfg}) at
high energies. Nuclei where very little of the discrete spectrum is known. In
these
cases,  (\ref{eq:gexp}) is used for the low- and intermediate-
excitation
portions of the spectra and (\ref{eq:gfg}) for the highly excited
part.
All three groups are spread across the whole periodic table.

\begin{wrapfigure}{r}{9cm}
\centering
\includegraphics[scale=0.1,angle=0]{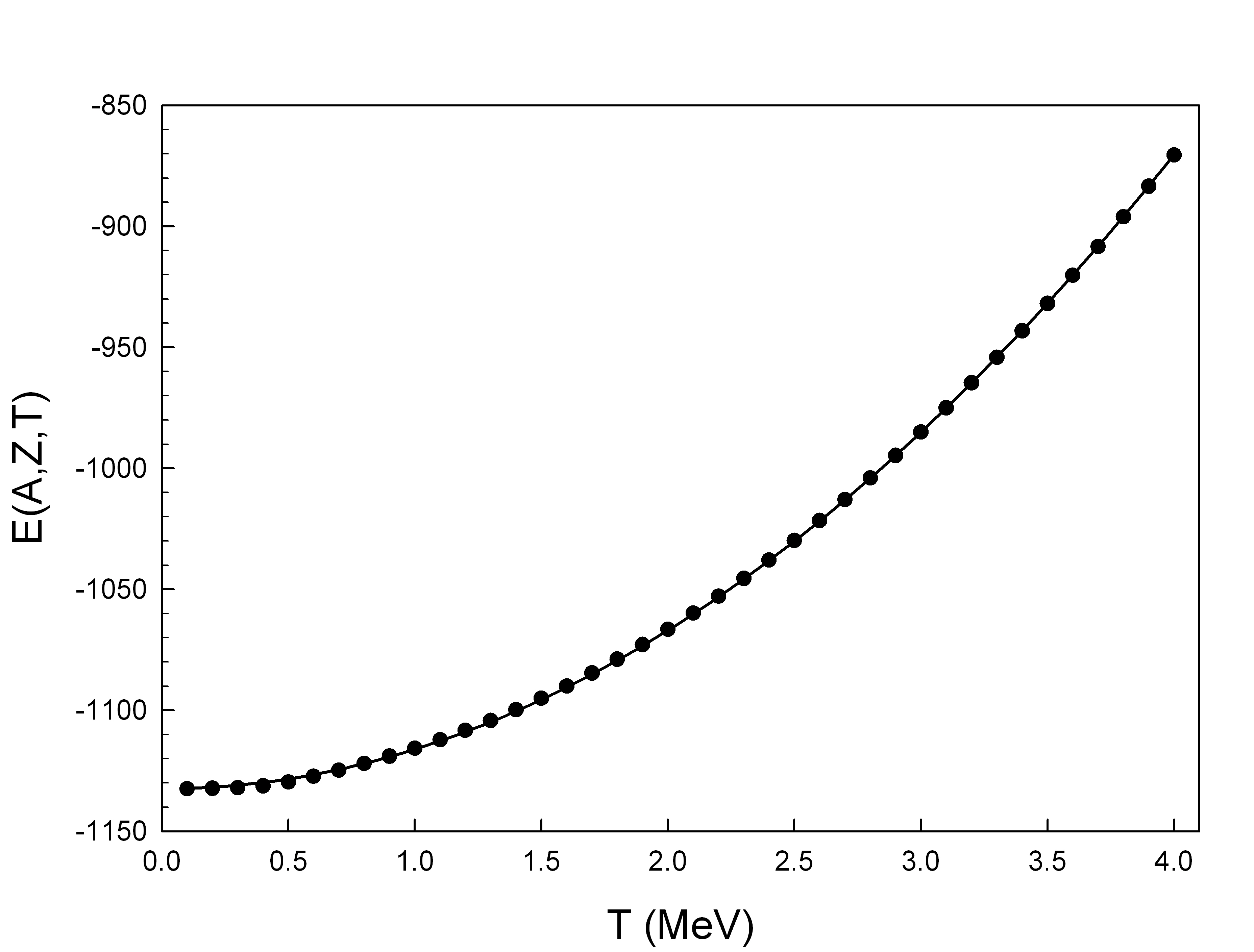} \center\vspace{-0.5cm}
\caption{\footnotesize{Quadratic fit of the calculated binding energy of Xe
($A$=130) using equation (\ref{binding energy}).  The fit to the Xe data is
representative of the fits obtained for other nuclei.}} \label{fig1}
\end{wrapfigure}

The lower bound $E_n$ on the integral in (\ref{eq:part}) is taken to
be the
energy at which (\ref{eq:gfg}) should become valid (from \cite{GC65})
for the first case above, 80\% of the largest discrete energy level for the
second case
above, and
zero for the third case.  For temperatures up
to $T
\approx 4$ MeV,  the upper bound $E_{max} \approx 3$ GeV was used.

The coefficients in the mass formula have been determined by a least
squares fit of (\ref{eq:exc1}) to the ensemble average of the
excitation energy
\begin{equation}
\mathcal{E}(A,Z,T) = - \frac {\partial } {\partial {\beta}}
\ln\mathcal{Z}(A,Z,T)
\label{eq:eex}
\end{equation}
determined from a total of 313 nuclei in the mass region  $22 \leq A
\leq
250$ for
temperatures $T \leq 4$ MeV. The temperature dependence of the six
coefficients is given in reference \cite{DHMMT93}.
 This finite temperature
parametrization has been used to identify the pairing phase transition in
symmetric nuclear matter \cite{RMK01}. The same techniques have also been
successfully employed  to identify the remnants of the pairing phase transition
in finite nuclei \cite{KMPPF05}.

\begin{figure}
\begin{minipage}[b]{.46\linewidth}
\begin{center}
\includegraphics[scale=0.12,angle=0]{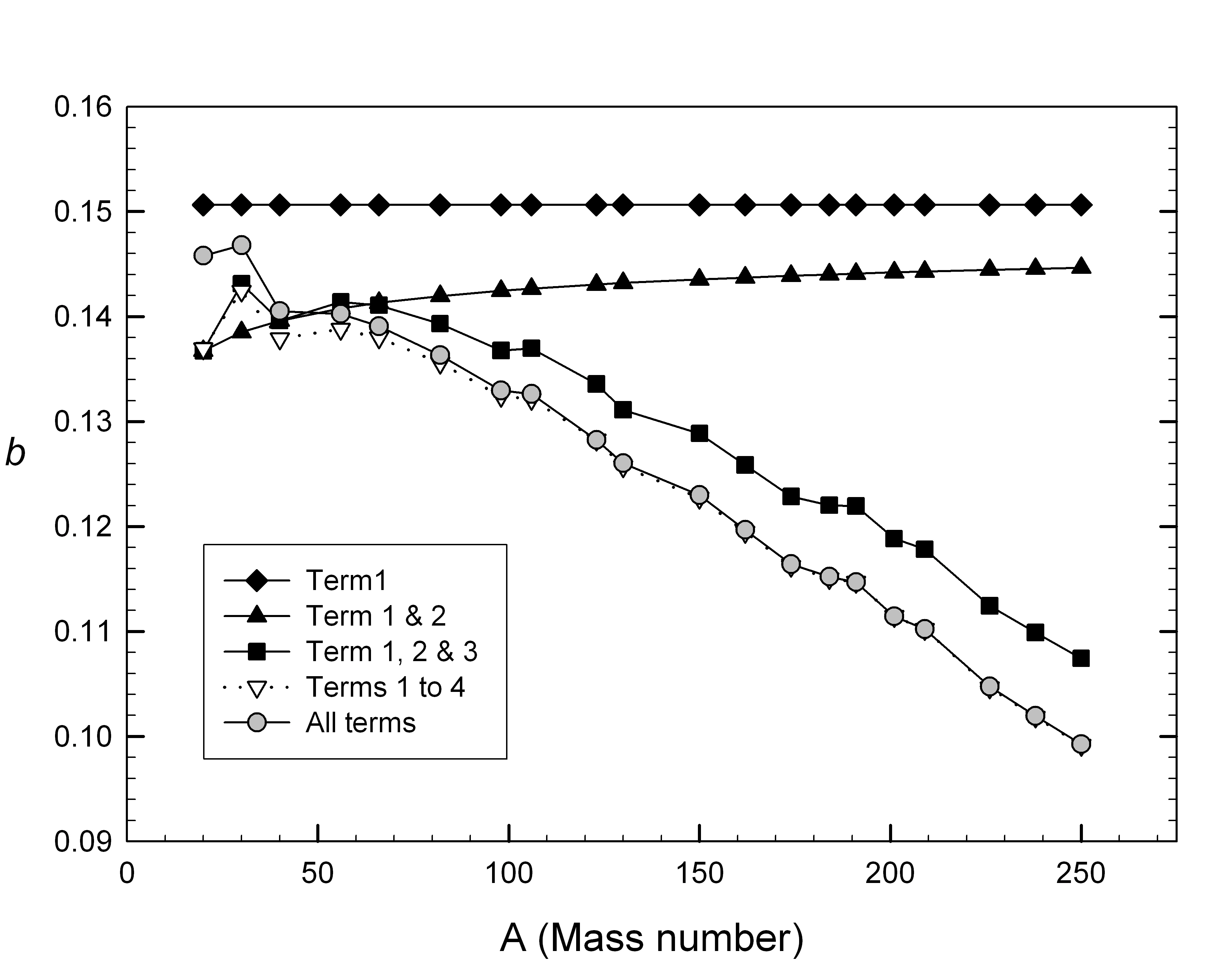}\center\vspace{-0.5cm}
\caption{\footnotesize{The resultant $b$ from various combinations of terms in
equation (\ref{binding energy}) has been plotted against $A$, thereby
illustrating the linearity of all five cases.}} \label{fig2}
\end{center}
\end{minipage}\hfill
\begin{minipage}[b]{.46\linewidth}
\begin{center}
\includegraphics[scale=0.12,angle=0]{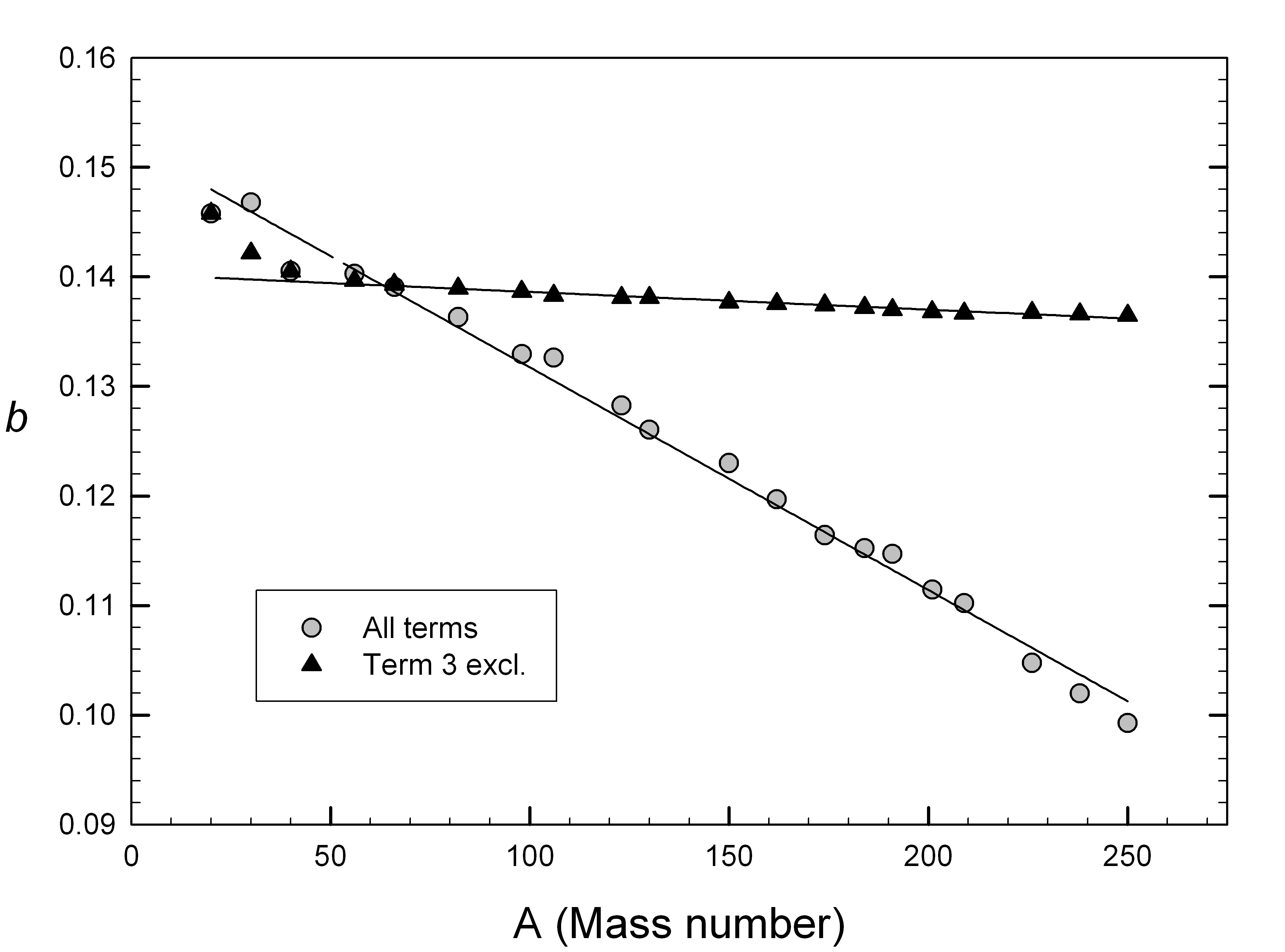}\center\vspace{-0.5cm}
\caption{\footnotesize{Linear fit of the $b$ resulting from equation
(\ref{binding energy}) and equation (\ref{binding energy}) without term 3 has
been plotted.\\}} \label{fig3}
\end{center}
\end{minipage}
\end{figure}

The excitation energy, $\mathcal{E}(A,Z,T)$, has been calculated for a number
of stable isotopes in the mass region $20\leq A\leq 250$ using the temperature
dependent coefficients determined in reference \cite{DHMMT93}. For each stable
isotope, the constant $b$ in equation (\ref{et}) was determined from a quadratic
fit of $\mathcal{E}(A,Z,T)$ versus $T$. These fits were in general excellent, as
shown in figure \ref{fig1} with only occasional small deviations from the
quadratic fit occurring at low temperature. This is not surprising given that
the continuum contributions are given by Fermi gas expressions. 
However, this alone in no way guarantees any simple dependence of $b$ on $A$.
Note also  that
no attempt has been made in the fits  to take into account the presence of low
temperature collective to non-collective phase
transitions \cite{MCQ89,KMPPF05}. The
coefficient $b$ was then plotted as a function of $A$, as shown in figure
\ref{fig2}.
As can be seen in figures \ref{fig2} and \ref{fig3} there is a relatively
simple dependence of $b$ on $A$. 
In order to determine the effect on $b$ of the different terms in the
temperature dependent binding energy (\ref{binding energy}), various
combinations of the five terms were plotted against $T$ then quadratically
fitted to obtain $b$ for each $A$ and finally these points were linearly fitted,
as shown in figures \ref{fig2} and \ref{fig3}.  Although only 20 stable nuclei
were used these were chosen at random to cover the range from $A$=20 to $A$=250.
 In figure \ref{fig3} the equations for the fits are
$b=-2.032\times10^{-4}A+0.152$ when all terms of equation (\ref{binding energy})
are used and $b=-1.617\times10^{-5}A+0.140$ if term 3 of equation (\ref{binding
energy}) is excluded. The gradient is small in both cases but especially small
when term 3 is excluded, with $b$ gradually varying from 0.141 to 0.137
for nuclei larger than $A$=40. Surface effects become significant in
nuclei smaller than $A$=40 and the pairing effects are most likely
overestimated. This
most probably accounts for the deviation of the corresponding points from the
fit in figure \ref{fig3} and subsequently prompted the exclusion of these points
from the linear fit.  These results demonstrate that the asymmetry term ($i.e.$
term 3) is largely responsible for the $A$ dependence of $b$. 

Furthermore in the Canonical ensemble the specific heat is given by
\begin{eqnarray}
 C&=&\frac{\partial \mathcal{E}}{\partial T} \\
  &=& T\frac{\partial S}{\partial T}. 
\end{eqnarray}
At low $T$ up to an additive constant the entropy per particle, $S$, for finite
nuclei in
the mass range $20\leq A\leq 250$ is therefore given simply by 
\begin{equation}
 S=\frac{1}{2} b T
\end{equation}
with $b=-2.032\times10^{-4}A+0.152.$ 

It is interesting to note that the simple Fermi gas model does not take
properly into account that the nucleus is composed of protons and neutrons and 
that therefore the effects of the Coulomb force must be considered.  Were only
the nuclear force present one might expect $b$ to be independent of $A$. The
presence of the Coulomb force gives rise to the asymmetry term which at $T$=0
leads to a displacement of the island of stability away from nuclei with $N=Z$ and
at finite temperature is largely responsible for an $A$ dependence in $b$.
 
\end{document}